\documentclass[a4paper,11pt]{article}
\usepackage{jheppub} 
\usepackage{xspace}
\usepackage{multicol}

\newcommand{\met}{\ensuremath{\hbox to 0pt{$\not$}E_T}\xspace}

\title{\boldmath  Probing a left-right symmetric model from a displaced shower at the CMS muon system}

\author[a]{Wei Liu,}
\author[a,b]{Zixiang Chen,}
\affiliation[a]{Department of Applied Physics and MIIT Key Laboratory of Semiconductor Microstructure and Quantum Sensing, Nanjing University of Science and Technology, Nanjing 210094, China}
\affiliation[b]{University College London, Gower Street, London WC1E 6BT, United Kingdom}
\emailAdd{wei.liu@njust.edu.cn}

\abstract{
We study the potential collider discovery for the right-handed neutrinos~(RHNs) in the minimal left-right symmetric model~(LRSM). In this model, the RHNs can be either produced from the right-handed gauge boson $W_R^\pm$, or the scalar triplet $\Delta$ decays. The RHNs can subsequently decay via the heavy $W_R^\pm$, making them potentially long-lived.
In looking for such long-lived RHNs, a search
for displaced shower at CMS muon system is reinterpreted for either the processes, $pp \rightarrow W_R^\pm \rightarrow \ell^\pm N$ or $pp \rightarrow \Delta \rightarrow N N$. From the former process, we exclude $M_{W_R} \lesssim 3.6$ TeV using current CMS data,
or $\lesssim 8.0$ TeV at the future HL-LHC. For the $pp \rightarrow \Delta \rightarrow N N$ process, HL-LHC is able to probe much larger parameter space as long as the proper decay length of the RHNs, 0.01 m $ \lesssim L_0 \lesssim$
1 m. This implies that the displaced shower searches are able to explore the parameter space of the LRSM which have not been excluded yet. 
}

\begin{document}
\maketitle
\flushbottom

\section{Introduction}
\label{sec:intro}
The origin of neutrino mass remains one of the most fundamental open questions in particle physics. The standard model (SM) lacks a natural explanation for the tiny but non-zero neutrino masses, which were first observed in neutrino oscillation experiments\cite{Davis:1994jw, Super-Kamiokande:1998kpq, KamLAND:2002uet,SNO:2002tuh}. One of the most elegant and widely studied explanations is the seesaw mechanism~\cite{Minkowski:1977sc,Mohapatra:1980yp,Schechter:1980gr}, which provides a natural approach to generating small neutrino masses by introducing right-handed neutrinos (RHNs). Among the various theoretical frameworks that incorporate the seesaw mechanism, the minimal left-right symmetric model (LRSM) is a well-motivated extension of the Standard Model that restores parity at high energies~\cite{Pati:1974yy,Mohapatra:1974hk,Mohapatra:1974gc,Senjanovic:1975rk,Senjanovic:1978ev,Mohapatra:1979ia}.

The LRSM can lead to both type-I and II seesaw mechanisms that naturally give small masses to SM neutrinos. It also predicts other particles not included in the SM, they are heavy gauge bosons $W_R^\pm$ and $Z^\prime$, as well as the scalar triplet $\Delta$, extending the SM gauge group to $SU(2)_L \times SU(2)_R \times U(1)_{B-L}$, thereby restoring left-right symmetry at high energy scales. 
The main search channel for the LRSM at the Large Hadron Collider (LHC) is the Keung-Senjanović (KS) process~\cite{Keung:1983uu}, which involves the production of a heavy charged gauge boson $W_R^\pm$, which subsequently decays into a lepton and a RHN, finally
leading to a pair of same-sign leptons plus jets. 
Despite extensive searches for these prompt final states by the CMS~\cite{CMS:2012zv,CMS:2016fxb,CMS:2017xcw,CMS:2018iye,CMS:2018agk,CMS:2021dzb} and ATLAS~\cite{ATLAS:2015gtp,ATLAS:2018dcj,ATLAS:2019isd,ATLAS:2023cjo}, no signal has been observed yet, excluding $M_{W_R} \lesssim$ 4.0 TeV~\cite{ATLAS:2019fgd}. The RHNs can also be long-lived, given that they are light and $W_R^\pm$ is heavy. Searches for long-lived RHNs from KS processes are proposed in Ref.~\cite{Nemevsek:2018bbt, Cottin:2018kmq, Cottin:2019drg, Urquia-Calderon:2023dkf}, performed both at the LHC as well as future lepton colliders, such as the FCC-ee~\cite{FCC:2018byv}, CEPC~\cite{CEPCStudyGroup:2018rmc}, ILC~\cite{ILC:2007oiw} and CLIC~\cite{Aicheler:2012bya}. 
In addition, the RHNs can be pair-produced from the decay of the scalar triplet $\Delta$. This is studied at the LHC~\cite{Maiezza:2015lza, Dev:2016dja, Nemevsek:2016enw}, as well as lepton colliders~\cite{Biswal:2017nfl}, for prompt final states, and long-lived final states. 

Most of the existing studies for long-lived RHNs take the displaced vertex as their potential signatures.
In addition to displaced vertex, displaced showers represent another distinct signature in the search for long-lived particles (LLPs)~\cite{CMS:2021juv}. A displaced shower occurs when a heavy particle decays after traveling a macroscopic distance, entering the volume of the CMS muon system. This decay produces a spread of particles, leading to a large number of hits, making clusters in the CMS muon system. The searches for them can be reinterpreted to other scenarios of LLPs, to extend their application. Such reinterpretation has already been done for the minimal RHNs model~\cite{Cottin:2022nwp}, as well as the $B-L$ model~\cite{Liu:2024fey}, and other models~\cite{Mitridate:2023tbj}.

In this work, we reinterpret the displaced shower searches as long-lived RHNs in the LRSM model.
The model predicts the existence of heavy RHNs, which can have a long lifetime. The crucial part is the right-handed gauge boson $W_R^\pm$, which mediates interactions involving the right-handed neutrinos. The mass and coupling structure of $W_R^\pm$ influence the production and decay properties of these RHNs. 
Furthermore, the triplet scalar $\Delta$ plays a fundamental role in neutrino mass generation~\cite{Mohapatra:1980yp}, opening another production channel of the RHNs. By analyzing displaced showers resulting from the long-lived RHNs decay in these processes, we can test the LRSM.  Specifically, we will focus on long-lived RHNs which can be produced via the processes $pp \rightarrow W_R^{\pm} \rightarrow l^{\pm} N$ and $gg \rightarrow \Delta \rightarrow N N$.

Compared to the existing literature Ref.~\cite{Nemevsek:2016enw},
our work focuses on long-lived signatures at the LHC in a more realistic setting. We recast the relatively new CMS displaced shower search, incorporating detector geometry, reconstruction, and trigger efficiencies. This allows us to set robust limits with current CMS data and provide reliable projections for the HL-LHC. Compared with prompt searches, which dominate current collider bounds, the displaced shower signal benefits from intrinsically low Standard Model backgrounds and hence provides a complementary avenue for discovery. We will show that displaced showers from both $pp \to W_R \to \ell N$ and $gg \to \Delta \to NN$ can be probed at the 13 TeV LHC and HL-LHC, extending the accessible parameter space well beyond existing limits.
By focusing on signatures that can be realistically studied at the LHC with existing data and near-future upgrades, our work complements the theoretical projections of Ref.~\cite{Nemevsek:2016enw} and Ref.~\cite{Biswal:2017nfl}, and provides immediate input to the ongoing LHC long-lived particle program.

The structure of the paper is as follows. Section \ref{model} introduces the theoretical framework of the minimal LRSM model, including its gauge structure, scalar pattern, and particle content. We also analyze the primary production mechanisms of RHNs and discuss their decay modes. Section \ref{dis} explores the possibility of detecting long-lived RHNs through displaced showers in the CMS muon system, highlighting the efficiency of this search as a function of the model parameters. Section \ref{sen}, we present a sensitivity study for two key channels: \( pp \rightarrow W_R^{\pm} \rightarrow l^{\pm} N \) and \( gg \rightarrow \Delta \rightarrow N N \), assessing the reach of current and future runs of the LHC. Finally, we conclude in section \ref{con}.

\section{Model}
\label{model}
\subsection{The Minimal Left-Right Symmetric Model}
The minimal left-right symmetric model is an extension of the Standard Model that restores parity symmetry at high energy scales~\cite{Mohapatra:1979ia}. In the SM, weak interactions violate parity symmetry, meaning that they do not behave in the same way when left and right are swapped. The LRSM addresses this by introducing right-handed counterparts to the left-handed particles found in the SM, thereby making the theory symmetric under left-right transformations. Turning the tables around, the LRSM extends the SM guage group $ \mathrm{SU}(2)_L \times \mathrm{U}(1)_{Y}$ to $\mathrm{SU}(2)_L \times \mathrm{SU}(2)_R \times \mathrm{U}(1)_{B-L}$, where $B-L$ represents the difference between the baryon number ($B$) and the lepton number ($L$). 

This extension introduces an additional ${SU}(2)_R$ gauge group that brings new gauge bosons: charged gauge bosons denoted as $W_R^\pm$ and a neutral gauge boson denoted as $Z^{\prime}$, distinct from the SM $W$ and $Z$ bosons. By treating left-handed and right-handed particles symmetrically, the LRSM restores parity symmetry at high energy scales.
The framework incorporates a discrete generalized charge conjugation $\mathcal C$ or parity conjugation $\mathcal P$, which restricts the number of free parameters by setting the gauge couplings of the left- and right-handed groups ${SU}(2)$ equal at tree level, i.e. $g_L=g_R$~\cite{Urquia-Calderon:2023dkf}.

In this set-up,  in the scalar sector, 
two triplets $\Delta_{L, R}$ is introduced~\cite{Urquia-Calderon:2023dkf}, 
\begin{equation}
\Delta_{L, R}=\left(\begin{array}{cc}
\frac{\Delta_{L, R}^{+}}{\sqrt{2}} & \Delta_{L, R}^{++} \\
\Delta_{L, R}^0 & -\frac{\Delta_{L, R}^{+}}{\sqrt{2}}
\end{array}\right).
\end{equation}
These scalars acquire the following vacuum expectation values~(VEV),

\begin{equation}
\left\langle\Delta_{L, R}^0\right\rangle=\frac{v_{L, R}}{\sqrt{2}},
\end{equation}
where $v_R$ is the VEV which is predominantly responsible for the breaking of ${SU(2)_R} \times U(1)_{B-L}$ into $U(1)_Y$ group. The right-handed gauge bosons thus acquire masses as,
\begin{equation}
\begin{aligned}
M_{W_R}^2 \simeq \frac{1}{2} g^2 v_R^2, \quad M_{Z^{\prime}}^2 \simeq g^2 v_R^2 \frac{c_w^2}{c_{2 w}}, \quad 
M_{Z^{\prime}} \simeq 1.69 M_{W_R}, 
\end{aligned}
\end{equation}
where $g\equiv g_L \simeq g_R$, and $c_w \equiv \cos \theta_w$ is the cosine of the Weinberg angle. 

The presence of $\Delta_{L}$ and the VEV $v_L$ is responsible for generating the tiny neutrino masses via a type-II seesaw~\cite{Mohapatra:1980yp,Magg:1980ut,Schechter:1980gr,Cheng:1980qt,Lazarides:1980nt}, which imposes $v_L \ll v, v_R$, with $v$ being the SM VEV~\cite{Dev:2018sel}.

In order to further break $SU(2)_L \times U(1)_Y$ into $U(1)_{EM}$, the $U(1)$ group for the electro-weak~(EW), a bidoublet $\Phi$ is added,
\begin{equation}
\quad \Phi=\left(\begin{array}{cc}
\phi_1^0 & \phi_2^{+} \\
\phi_1^{-} & -\phi_2^{0 *}
\end{array}\right),
\end{equation}
with 
\begin{equation}
\quad\langle\Phi\rangle=\frac{1}{\sqrt{2}} \operatorname{diag}\left(v_1, -v_2 e^{-i \alpha}\right),
\end{equation}
where $v^2 = v_1^2 + v_2^2$. 

In the mass eigenstates of the neutral scalars, only the SM-like Higgs $h$ and the triplet-like $\Delta$ are involved~\cite{Nemevsek:2016enw}, 
\begin{align}
  \begin{pmatrix} h \\ \Delta \end{pmatrix} = \begin{pmatrix} c_\theta & s_\theta \\ -s_\theta & c_\theta \end{pmatrix} 
  \begin{pmatrix} 
    h_0 \\ \Delta_0 \end{pmatrix},
\end{align}
where $h_0 = \text{re } \phi_1^0 \, \left(\frac{v_1}{v} \right) + \text{re } \phi_2^0 \, \left(\frac{v_2}{v} \right)$, $s_\theta = \sin \theta$, $c_\theta = \cos \theta$ is the scalar mixing angle. $\Delta_0$ is the real part of $\Delta_R^0$.
The behavior of $\Delta$ is similar to that of a SM singlet in a good approximation. Hence, the current limits on the scalar mixing are roughly $s_\theta \lesssim 0.24$ for $m_\Delta > m_h$ from the signal rates measurements at the LHC~\cite{Robens:2022cun, Papaefstathiou:2022oyi, ATLAS:2021vrm}, and $s_\theta \lesssim 0.1$ for 10 GeV $ \lesssim m_\Delta \lesssim$ 100 GeV from LEP searches for light singlet-like scalars~\cite{Robens:2015gla,Ellis:2022lft}. We therefore take $s_\theta = 0.1$ as our benchmark in the rest of the paper.

In order to further break $SU(2)_L \times U(1)_Y$ into $U(1)_{EM}$, the $U(1)$ group for the electro-weak~(EW), a bidoublet $\Phi$ is added,
\begin{equation}
    \Phi = \begin{pmatrix}
\phi_1^0 + v_1 & \phi_2^+ \\
\phi_1^- & \phi_2^0 + v_2 e^{i\alpha}
\end{pmatrix},
\end{equation}
with
\begin{equation}
\langle\Phi\rangle = \frac{1}{\sqrt{2}} \operatorname{diag}\left(v_1, -v_2 e^{-i \alpha}\right).
\end{equation}
The neutral scalars in the bidoublet $\Phi$ can mix with the real component of $\Delta_R^0$, denoted $\Delta_0$, through terms in the scalar potential. Mixings involving $\Delta_L$ are suppressed due to the smallness of $\langle\Delta_L\rangle \propto v^2/v_R$, and those with the flavor-violating scalar $H$ are phenomenologically constrained~\cite{Maiezza:2016bzp}. Therefore, in the mass eigenstates of the neutral scalars, only the SM-like Higgs $h$ and the triplet-like $\Delta$ are involved~\cite{Nemevsek:2016enw}, 
\begin{align}
  \begin{pmatrix} h \\ \Delta \end{pmatrix} = \begin{pmatrix} c_\theta & s_\theta \\ -s_\theta & c_\theta \end{pmatrix} 
  \begin{pmatrix} 
    h_0 \\ \Delta_0 \end{pmatrix},
\end{align}
where $h_0 = \text{re } \phi_1^0 \, \left(\frac{v_1}{v} \right) + \text{re } \phi_2^0 \, \left(\frac{v_2}{v} \right)$, $s_\theta = \sin \theta$, $c_\theta = \cos \theta$ is the scalar mixing angle. $\Delta_0$ is the real part of $\Delta_R^0$.
The behavior of $\Delta$ is similar to that of a SM singlet in a good approximation. Hence, the current limits on the scalar mixing are roughly $s_\theta \lesssim 0.24$ for $m_\Delta > m_h$ from the signal rates measurements at the LHC~\cite{Robens:2022cun, Papaefstathiou:2022oyi, ATLAS:2021vrm}, and $s_\theta \lesssim 0.1$ for 10 GeV $ \lesssim m_\Delta \lesssim$ 100 GeV from LEP searches for light singlet-like scalars~\cite{Robens:2015gla,Ellis:2022lft}. We therefore take $s_\theta = 0.1$ as our benchmark in the rest of the paper.

For the fermion sector, in the interaction basis, the left-handed and right-handed fermion fields are arranged in doublets under $\mathrm{SU}(2)_L$ and $\mathrm{SU}(2)_R$ respectively,
\begin{equation}
Q_{L, R}=\binom{u^\prime}{d^\prime}_{L, R}, \quad L_{L, R}=\binom{\nu^\prime}{\ell^\prime}_{L, R} .
\end{equation}

Through a type I $+$ II seesaw, neutrinos mass terms are introduced~\cite{Kriewald:2024cgr}
\begin{align}
\begin{split}
  &\mathcal L_{\nu\text{-mass}} = -\frac{1}{2} \left(\bar\nu_L^\prime\:\bar\nu_R^{\prime c}\right)
  \begin{pmatrix}
    M_L & M_D 
    \\
    M_D^T & M_R
  \end{pmatrix}
  \begin{pmatrix} \nu_L^{\prime c} \\ \nu_R^\prime \end{pmatrix} + \text{H.c.} \, ,
\end{split}  
\end{align}
where $M_L = v_L Y_L^M$, $M_D = Y_\ell v_1 - \tilde Y_\ell e^{- i \alpha} v_2 $
and $M_R = v_R Y_R^M$.
Up to leading order in $M_R^{-1}$,
\begin{align} \label{eqn:seesaw}
  M_\nu &\simeq M_L - M_D M_R^{-1} M_D^T \, ,\quad M_N \simeq M_R .
\end{align}
And they can be diagonalized into $\mathrm{diag}(m_{\nu_1}, m_{\nu_2}, m_{\nu_3})$ and $\mathrm{diag}(m_{N_1}, m_{N_2}, m_{N_3})$ via unitary rotations. For simplicity, among the three RHNs, we assume that only one of them is accessible at colliders.

The main interactions of the RHNs contain charge-current interactions with $W_R^\pm$,
\begin{eqnarray}
  \mathcal L_{cc}^\ell = 
  \frac{g}{\sqrt{2}} \bar\ell_R \gamma^\mu \mathcal U_R N W_R^\mu \, .
\end{eqnarray}
where $\mathcal U_R$ is a semi-unitary matrix.
Besides, they also interact with $\Delta$ via Yukawa terms,
\begin{align}
  \begin{split}
  \mathcal L_Y^\ell &= 
  \bar L_R^{\prime c} i \sigma_2 \Delta_R Y_R^M L_R^\prime + \text{H.c.} \, .
  \end{split}
\end{align}
The RHNs can further couple to the light neutrinos and Higgs via the Dirac mass terms. However, it is neglected here since the Dirac Yukawa should be small to explain the tiny neutrino masses. The interactions induced by the active-sterile mixing are also neglected for similar reasons.
Hence, in the rest of the paper, we take $N$ to be predominantly coupled to $W_R^\pm$ and also $h/\Delta$.

\subsection{Production and Decays of the RHNs at Colliders}
\begin{figure}[t!]
    \centering
    \includegraphics[width=0.50\textwidth]{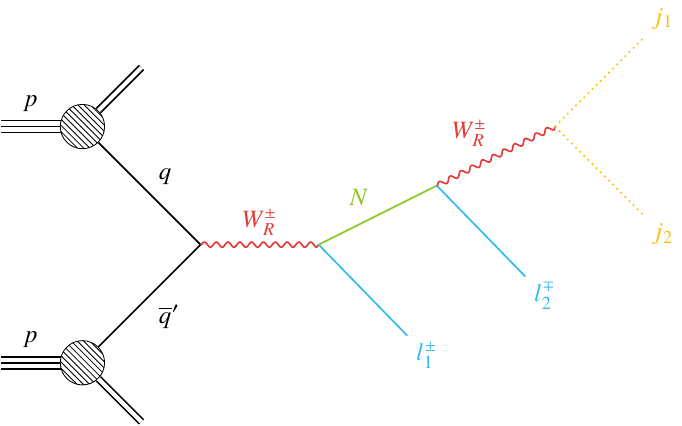}
    \includegraphics[width=0.43\textwidth]{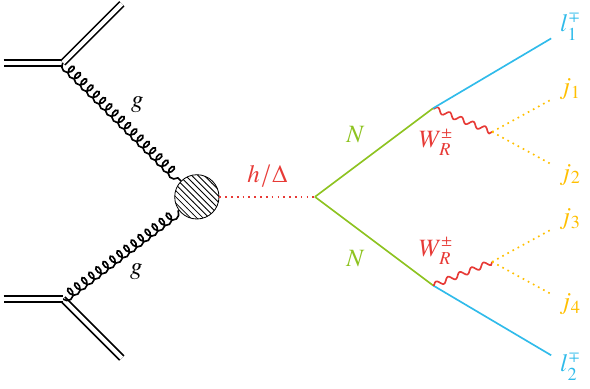}
    \caption{The Feynman diagrams of the processes $pp \rightarrow W_R^{\pm} \rightarrow l^{\pm} N$ and $gg \rightarrow h/\Delta \rightarrow N N$. }
    \label{fig:feyn}
\end{figure}

Based on the interactions, there are three production modes to produce $N$ at the LHC~\footnote{Another process $pp \rightarrow h \rightarrow \Delta \Delta \rightarrow 4 N$ is also considered in Ref.~\cite{Nemevsek:2016enw}.},
\begin{itemize}
    \item $pp \rightarrow W_R^{\pm} \rightarrow l^{\pm} N$,
    \item $gg \rightarrow h/\Delta \rightarrow N N$.
\end{itemize}
The Feynman diagrams are illustrated in Fig.~\ref{fig:feyn}.

\begin{figure}[t!]
    \centering
    \includegraphics[width=0.49\textwidth]{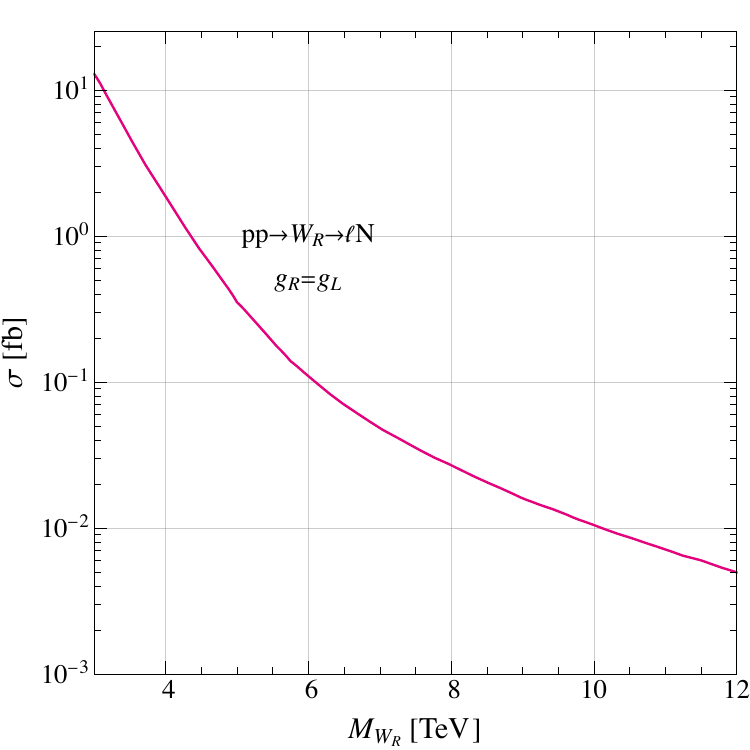}\includegraphics[width=0.49\textwidth]{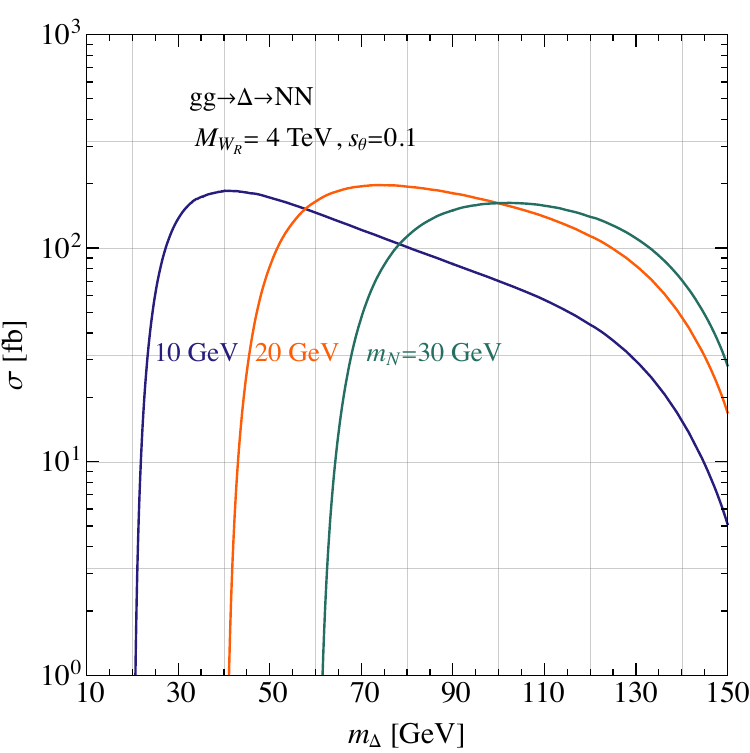}
    \caption{Left: The cross section of $pp \rightarrow W_R^{\pm} \rightarrow l^{\pm} N$ at the 13 TeV LHC. Right: the same for 
    $gg \rightarrow \Delta \rightarrow N N$, for fixed $m_N = 10, 20, 30$ GeV, $M_{W_R} = 4$ TeV and scalar mixing $s_\theta =$ 0.1.}
    \label{fig:csd}
\end{figure}

$pp \rightarrow W_R^{\pm} \rightarrow l^{\pm} N$ is the KS process. The production cross section of this process at the LHC is well studied; see, e.g.~\cite{Nemevsek:2018bbt}. Current best limits from CMS/ATLAS searches for dijets final states of $W_R^\pm$ put $M_{W_R} \gtrsim 4.0$ TeV~\cite{ATLAS:2019fgd}. With such a heavy $W_R^\pm$, $\sigma(pp \rightarrow W_R^{\pm} \rightarrow l^{\pm} N) \lesssim $ 10 fb at 13 TeV LHC, as shown in Fig.~\ref{fig:csd} left.

$pp \rightarrow h/\Delta \rightarrow N N$ come from the scalar mixing and the neutrino mass terms. 
The production cross section of the $gg \rightarrow h \rightarrow N N$ process can be expressed as 
\begin{align} \label{eqGamDhNN}
    \sigma(gg \rightarrow h \rightarrow N N)= \sigma(gg\rightarrow h) \times Br(h \rightarrow N N),
\end{align}
with $Br(h \rightarrow N N) = \Gamma(h\rightarrow N N)/\Gamma(h)$, and
\begin{align}
\Gamma(h \to N N) = s_\theta^2 \frac{\alpha_w}{8} m_h \left(\frac{m_N}{M_{W_R}} \right)^2 
  \beta_{h N}^{3/2},
\end{align}
where $\beta_{i N} = 1-(2 m_N/m_i)^2$ and $\alpha_w = g^2/(4\pi)$.

Therefore, at the 13 TeV LHC
\begin{align}
\sigma(gg \rightarrow h \rightarrow N N) \approx 0.1~\text{fb} \times \left(\frac{s_\theta}{0.1}\right)^2 \times \left(\frac{m_N}{10~\text{GeV}}\right)^2 \times \left(\frac{4~\text{TeV}}{M_{W_R}}\right)^2.
\end{align}

For the process of $gg \rightarrow \Delta \rightarrow N N$,
\begin{align} \label{eqGamDhNN}
    \sigma(gg \rightarrow \Delta \rightarrow N N)\approx s_\theta^2 \times \sigma(gg\rightarrow h(m_\Delta)) \times Br(\Delta \rightarrow N N).
\end{align}
As $\Delta$ roughly behaves as an SM singlet, $\sigma(gg\rightarrow h(m_\Delta))$ can be obtained from Ref.~\cite{Anastasiou:2016hlm}.
Here we restrict ourselves to $m_{\Delta} < 2 m_W$, since the heavier $\Delta$ will decay into gauge bosons dominantly. For such $\Delta$, the branching ratio
\begin{align}
Br(\Delta \rightarrow N N)\approx\Gamma_{\Delta \to N N}/(\Gamma(\Delta \to N N)+\Sigma \Gamma(\Delta \to f \overline f)),
\end{align}
where
\begin{align} \label{eqGamDNN}
  \Gamma_{\Delta \to N N} &= c_\theta^2 \frac{\alpha_w}{8} m_\Delta \left(\frac{m_N}{M_{W_R}} \right)^2 
  \beta_{\Delta N}^{3/2},
\end{align}
and
\begin{equation}
  \Gamma_{\Delta \to f \overline f} = s_\theta^2 \, \Gamma_{h \to f \overline f} \left(m_h \to m_\Delta, s_\theta = 0 \right).
\end{equation}

Using the above expressions, we show $\sigma(gg \rightarrow \Delta \rightarrow N N)$ in Fig.~\ref{fig:csd} right~\footnote{Our results roughly agree with Ref.~\cite{Nemevsek:2016enw}, but the peak value of the cross section is roughly 50\% smaller, might due to the missing of NLO effects~\cite{Maiezza:2015lza}.}. 
 For light $\Delta$, $\sigma(gg \rightarrow \Delta \rightarrow N N)$ can easily exceed $10$ fb, making it a promising mode. Since the production of $pp \rightarrow h \rightarrow N N$ is much smaller, among the two processes, we focus only on $gg \rightarrow \Delta \rightarrow N N$ in the following discussions.

For a quantitative perspective, taking $s_\theta = 0.1$, $m_N = 10$~GeV, and $m_\Delta = 30$~GeV as shown in Fig.~\ref{fig:csd} (right), the cross section reaches $\sigma(gg \to \Delta \to NN) \approx 10^2~\text{fb}$. At an integrated luminosity of $\mathcal{L} = 3000~\text{fb}^{-1}$, this yields approximately $N \approx 3 \times 10^5,$ which indicates that, despite the relatively low scale $M_{W_R} \sim 4$~TeV, the benchmark parameter region can still lead to sizeable event yields at the 13 TeV LHC.

The RHNs mainly decays into three body final states via $W_R^\pm$, and its decay width reads~\cite{Urquia-Calderon:2023dkf}, 
\begin{equation} \label{eq:GN}
  \Gamma_N \approx \frac{\alpha_w^2 m_N^5}{64 \pi M_{W_R}^4}
  \sum_{u,c; d,s} \left| V_{ud}^{\text{\tiny CKM}} \right|^2,
\end{equation}
leading to macroscopic decay length, 
\begin{equation} \label{eq:GN}
    L_N \simeq 4.2~\text{cm} \times \left( \frac{M_{W_R}}{4~\text{TeV}}\right)^4 \times \left( \frac{10~\text{GeV}}{m_N}\right)^5.
\end{equation}
Such decay length can lead to abundant interesting phenomenology at the LHC, including displaced vertex~\cite{Nemevsek:2018bbt, Cottin:2019drg}. In the following section, we will discuss another potential signatures, the displaced shower signal at the muon system of the CMS~\cite{CMS:2021juv, CMS:2024bvl, CMS:2024ake}.

\section{Displaced Shower at CMS Muon system}
\label{dis}
The search for displaced shower is using CMS muon system as a calorimeter to look for the shower products from the LLP decays~\cite{CMS:2021juv, CMS:2024bvl}.  These showers will result in a large number of hits to form clusters, with $N_{\text{hits}} >$ 130, in the cathode strip chambers~(CSC) detector. This allows us to identify the signal from the background events. 

The original searches were looking for a pair of long-lived scalars from the Higgs decays~\cite{CMS:2021juv, CMS:2024bvl}, and RHNs in a minimal seesaw scenarios~\cite{CMS:2024ake}. Since in the LRSM model, the RHN can also be long-lived, and their decay products can be energetic to trigger subsequent shower processes, it is possible to recast the displaced shower search to look for RHNs in the LRSM model. To aid in reinterpretation, a map of the cluster efficiency as a function of the electromagetic as well as hadronic energy in certain region of the muon system is provided in HEPData~\cite{CMS:2021juv,hepdata.104408.v2}~\footnote{The search is updated using larger region of the CMS muon system as described in Ref.~\cite{CMS:2024bvl}, however without providing the updated efficiency in HEPData.}.

To recast the displaced shower search, there are several steps to follow. First, we generate the signal events using the event generator {\tt MadGraph5aMC@NLO}~v3.5.4\cite{Alwall:2014hca}, with
the Universal {\tt FeynRules} Output (UFO)~\cite{Degrande:2011ua} of the LRSM model provided in Ref.~\cite{Mattelaer:2016ynf}. The shower, hadronisation and jet matching are performed by {\tt PYTHIA}~v8.311 \cite{Bierlich:2022pfr}. Clustering of events is achieved by using {\tt FastJet}~v3.2.1 \cite{Cacciari:2011ma}. To simulate the detector efficiency of the CSC detector of the CMS muon system to identify LLP showers, we use {\tt Delphes}~v3.5.1 \cite{deFavereau:2013fsa}, with an updated card and new modules~\cite{delphes_pr, CMS:2021juv}.

Once the signal events are simulated, we apply the same selection criteria of the CMS search~\cite{CMS:2021juv}.

\begin{itemize}
    \item $ \met > 200~\text{GeV}$, large missing transverse energy~(MET) is required to trigger the event. If there are no objects in the tracker and calorimeter system, we simulate up to two initial radiated jet to trigger on.

    \item $ N(J) > 1$, at least 1 jet with transverse momentum $p_T >$ 50 GeV and pseudorapidity $|\eta| < 2.4)$. This is also used to trigger the event, since large MET mostly comes with hard jets.

    \item $N(\text{Cluster}) \geq 1$, at least one cluster is required in the CSC, which has $|\Delta\phi| < 0.75$ between their location and the MET, to identify which of the RHN decays.

\end{itemize}

After implementing the above selections, the total efficiency can be expressed, as
\begin{eqnarray}
\epsilon_{\rm tot} &\approx& 
 N_{\rm{LLP}} \times \epsilon_{\rm geo}\times \epsilon_{\rm reco} \times \epsilon_{\rm tri} ,\\
& \equiv & N^{\rm CSC}_{\rm{LLP}}\times  \epsilon_{\rm reco}\times \epsilon_{\rm tri} \equiv f_{\text{cluster}} \times \epsilon_{\rm tri},
\end{eqnarray}
where $N_{\text{LLP}}$ is the number of long-lived RHNs in the final states, and $N_{\text{LLP}} = 1~(2)$ for the $W_R^\pm~(\Delta)$ mediated processes, respectively. $N^{\rm CSC}_{\rm{LLP}}$ is the same, but requires the RHNs to decay inside the muon system.

\begin{figure}[t!]
    \centering
    \includegraphics[width=0.49\textwidth]{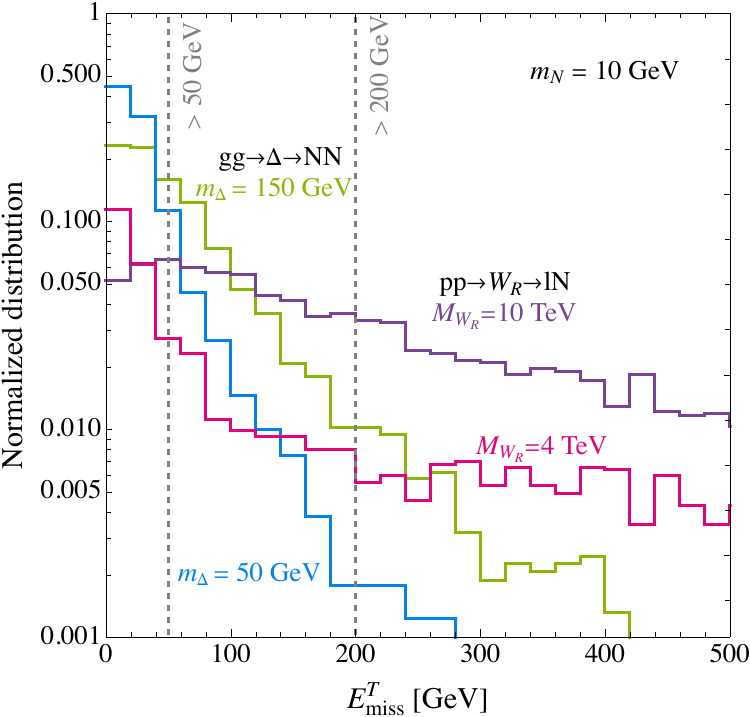}
    \caption{The missing transverse energy distribution for the process, $pp \rightarrow W_R^{\pm} \rightarrow l^{\pm} N$ with $M_{W_R} = 4$~TeV (red), 10 TeV~(purple). And for
    $gg \rightarrow \Delta \rightarrow N N$ with $m_{\Delta} = $ 50 GeV~(blue), 150 GeV~(green), while keeping $M_{W_R} = 4$~TeV. For all processes, we fix $m_N =$ 10 GeV.}
    \label{fig:MET}
\end{figure}

The trigger efficiency for requiring $\met > $ 200 GeV and $N(J)>$1 is encoded in
$\epsilon_{\rm tri}$. $N(J)>$1 can be satisfied by requiring initial radiated jets. The requirement $\met > $ 200 GeV is so powerful that only a few events can pass.
As shown in the Fig.~\ref{fig:MET}, for the process $pp \rightarrow \Delta \rightarrow N N$,  only less than 1~(5)\% of the events can pass the $\met >$ 200 GeV cut, for $m_{\Delta} = 50~(150)$ GeV.
In the contrary, with the help of the heavy parent $W_R^\pm$, more than 50\% of the events of the $pp \rightarrow W_R^\pm \rightarrow l^\pm N$ processes can pass this cut. Nevertheless, this cuts can be lowered to $ \met > 50~\text{GeV}$, which is made possible since a dedicated Level-1 and High Level Trigger for displaced shower is already installed and collecting data since LHC Run-3~\cite{Mitridate:2023tbj}, which we refer to as 'soft trigger'. With this soft trigger, all processes can have trigger efficiency $\gtrsim 50\%$, enabling one to observe many more events in future runs of the LHC.

The effects of requiring $N(\text{Cluster}) \geq 1$ are described by $f_{\text{cluster}}$, which means the fraction of the events satisfing this selection.
This requires that RHNs are sufficiently long-lived to decay within the volume of the CMS muon system, which roughly called for the decay length in the laboratory frame, $L_{\text{lab}} \equiv \beta \gamma L_0 \simeq 4-13$~m. However, these long-lived RHNs do not necessarily lead to a cluster. The clusters can be reconstructed from the final states of them, with efficiency $\epsilon_{\rm reco}$, provided in HEPData~\cite{CMS:2021juv,hepdata.104408.v2}, mainly as a function of the sum of the hadronic and electromagnetic energy of their decay products.

\begin{figure}[t!]
    \centering
    \includegraphics[width=0.49\textwidth]{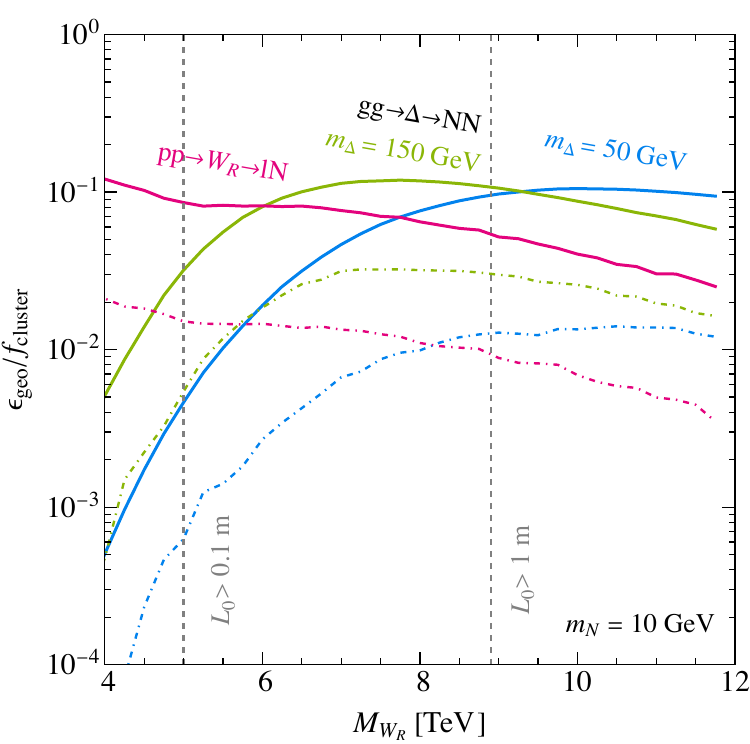}
   \includegraphics[width=0.49\textwidth]{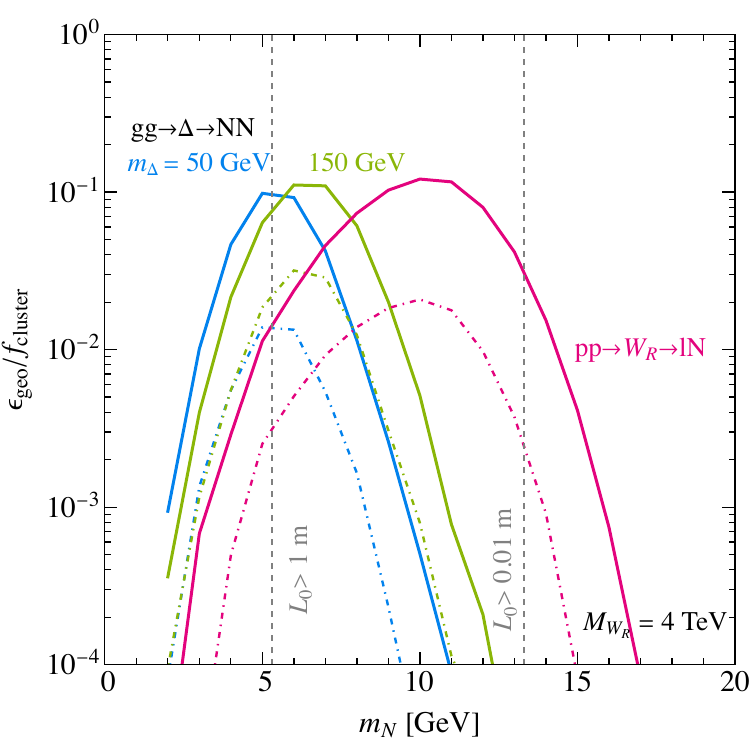}
    \caption{Left: Geometrical efficiency $\epsilon_{\text{geo}}$~(solid) and the cluster fraction $f_{\text{cluster}}$~(dashed solid) as a function of $M_{W_R}$ for the processes $pp \rightarrow W_R^{\pm} \rightarrow l^{\pm} N$~(red), $gg \rightarrow \Delta \rightarrow N N$ with $m_{\Delta} = $ 50 GeV~(blue), 150 GeV~(green). We fix $m_N =$ 10 GeV. Right: the same but for $m_N$ while fixing $M_{W_R} =$ 4 TeV.}
    \label{fig:dis}
\end{figure}

The above effects are described in Fig.~\ref{fig:dis}. In this figure, we show $\epsilon_{\text{geo}}$~(solid) and $f_{\text{cluster}}$~(dashed solid) as a function of $M_{W_R}$~(left) and $m_N$~(right), for the processes $pp \rightarrow W_R^{\pm} \rightarrow l^{\pm} N$ with $M_{W_R} = $ 4 TeV~(red), 10 TeV~(purple), and $gg \rightarrow \Delta \rightarrow N N$ with $m_{\Delta} = $ 50 GeV~(blue), 150 GeV~(green). The geometrical efficiency is controlled dominantly by the decay length in the laboratory frame, $L_{\text{lab}}$. It increases as $L_{\text{lab}}$ moves closer to the volume of the detector. 
For the $pp \rightarrow W_R^{\pm} \rightarrow l^{\pm} N$ process, since the RHNs are boosted by the remanent energy of the decaying heavy $W_R^\pm$, the Lorentz factor $\beta \gamma = p/m_N$ in this case can reach $\mathcal{O}(10^2)$. Hence, the efficiency peaks when $L_0 \simeq 0.01-0.1$ m. In this case, $L_{\text{lab}} \gtrsim 10$ m for $M_{W_R} \gtrsim 4$ TeV when $m_N =$ 10 GeV, and becomes even larger as $M_{W_R}$ increases from 4 TeV to 12 TeV. The same applies to decreasing $m_N$ from 10 GeV while keeping $M_{W_R} =$ 4 TeV.
Hence, $N$ is too long-lived, so the probability of it to decay within the detector decreases, reducing $\epsilon_{\text{geo}}$ from 0.1 to $\mathcal{O}(10^{-2})$.  When it comes to $gg \rightarrow \Delta \rightarrow N N$, the boost factor is only $\sim 10$, so the efficiency peaks when $M_{W_R} \simeq 7-10$ TeV, corresponding to $L_0 \simeq $ 1~m. Compared to geometrical efficiency, the reconstruction efficiency, however, is quite stable, as $\epsilon_{\text{recon}} \sim 0.1$ crosses different parameter spaces, resulting in the final cluster fraction $f_{\text{cluster}} \lesssim 10^{-2}$.

Besides the aforementioned cuts, we also apply the following cuts to further reduce background. We veto clusters which are too close to a jet, this is done by removing clusters which has $\Delta R = \sqrt{(\Delta\eta)^2 + (\Delta\phi)^2} < 0.4$ to the jet with $p_T >$ 10 GeV~. This removes background from LLPs inside the jet, e.g. $K_L$, or muon bremsstrahlung. We require
$-5~\text{ns} < \langle\Delta t_\text{CSC}\rangle < 12.5$~ns, average time of detector hits in the CSC cluster to the LHC clock is required to be small, to reject piled-up clusters. The effects of these cuts are however small for our signal processes.
As discussed, combing the effects from all selections, $\epsilon_{\text{tot}} \simeq \mathcal{O}(10^{-3})$ for the $W_R^\pm$ mediated processes, and $\simeq \mathcal{O}(10^{-4})$ for the $\Delta$ mediated ones.

Complying the above steps, we obtain the number of signal events, as
\begin{eqnarray}
N_{\text{signal}} = \mathcal{L} \times \sigma \times \epsilon_{\text{tot}},
\end{eqnarray}
where $\mathcal{L}$ represents the integrated luminosity, and $\sigma$ is the cross section of the signal processes. 

Given in Ref.~\cite{CMS:2021juv}, number of expected background $N_{\text{back}} = 2.0\pm 1.0$, for observed event $N = 3$. We exclude signal events $N_{\text{signal}} > 6.1$ at 95\% confidence level~(CL) with $\mathcal{L} = 137 $ fb$^{-1}$ at the 13 TeV LHC. For HL-LHC with $\mathcal{L} = 3000 $ fb$^{-1}$, we apply $ \met > 50~\text{GeV}$ instead, while the number of background events can still be negligible after requiring a larger cluster with more $N_{\text{hits}}$, which also reduces the number of signal events to 60\%~\cite{Cottin:2022nwp}. In this case, we require $N_{\text{signal}} \gtrsim 3.0$ at 95\% CL.

\section{Sensitivity}
\label{sen}
We will discuss the sensitivity on the parameter space ($M_{W_R}$, $m_N$), 
using the searches for displaced shower signature from the processes $pp \rightarrow W_R^{\pm} \rightarrow l^{\pm} N$, and $gg \rightarrow \Delta \rightarrow N N$ in this section.

\begin{figure}[t!]
    \centering
    \includegraphics[width=0.95\textwidth]{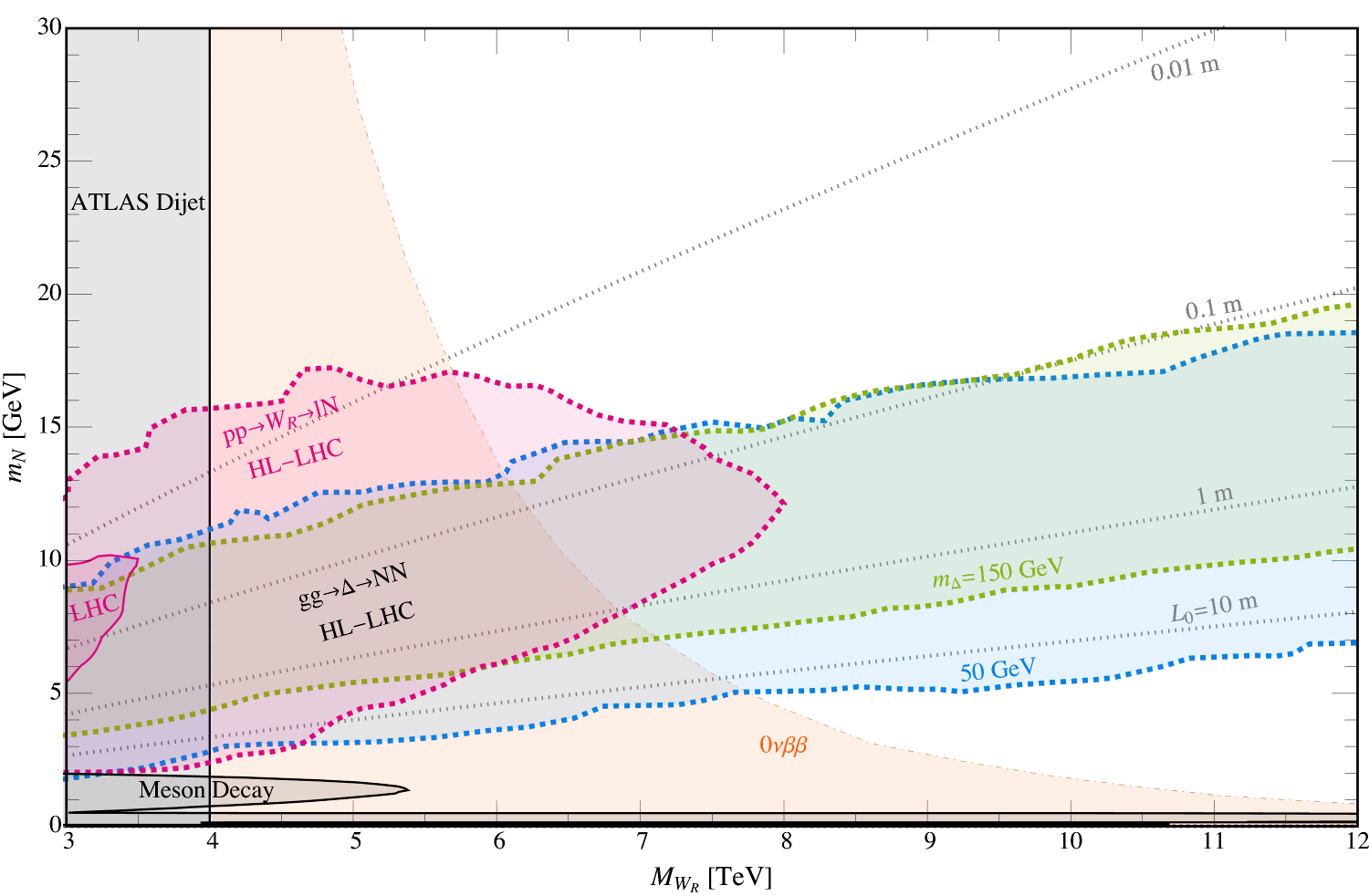}
    \caption{Sensitivity on the parameter space ($M_{W_R}$, $m_N$), 
using the searches for displaced shower signature from the processes $pp \rightarrow W_R^{\pm} \rightarrow l^{\pm} N$~(red), and $gg \rightarrow \Delta \rightarrow N N$ with $m_{\Delta}=$ 50 GeV~(blue) and 150 GeV~(green).  Current limits from dijets searches at the ATLAS~\cite{ATLAS:2019fgd}, as well as pseudoscalar meson leptonic decay data~\cite{Alves:2023znq} are shown for comparison. In addition, the limits from 0$\nu\beta\beta$ half life using existing Gerda-II data for a benchmark scenario where $t_\beta \equiv v_2/v_1 =$ 0.3 is also overlaid~\cite{Nemevsek:2023hwx}.}
    \label{fig:sen}
\end{figure}

Before that, we briefly summarize the existing limits.
These include direct searches for RHNs from $W_R^\pm$ decays~\cite{ATLAS:2023cjo, CMS:2021dzb}, and for dijets signals which can be interpreted as a jet pair from $W_R^\pm$ decays at the CMS and ATLAS~\cite{ATLAS:2019fgd, CMS:2019gwf}. For the direct searches, since only prompt final states are looked for, 
the limits only apply when $m_N \gtrsim $ 100 GeV, beyond the scope of our interests. For $m_N < $ 2 GeV, the RHNs from $W_R^\pm$ decays can be well constrained by the pseudoscalar meson leptonic decay data, excluding $M_{W_R} \lesssim 4-19$ TeV~\cite{Alves:2023znq}.
However, the limits from the dijets searches apply whatever $m_N$, excluding $M_{W_R} <$ 4.0 TeV~\cite{ATLAS:2019fgd}. Besides, there exist the limits from 0$\nu\beta\beta$ experiments. The LRSM can contribute to 0$\nu\beta\beta$ half life via the processes including the right-right heavy $N$ exchange and the left-right
amplitude via gauge boson mixing. The gauge boson mixing is controlled by $t_\beta \equiv v_2/v_1$, and we adopt the limits in Ref.~\cite{Nemevsek:2023hwx}, where a benchmark $t_\beta = 0.3$ is taken. Smaller $t_\beta$ can result in significantly worse limits, making $0 \nu \beta \beta$ experiments less competitive.

Reinterpreting the displaced shower signal as RHNs from $W_R^\pm$ decays, we present the first limits from long-lived final states using existing data in Fig.~\ref{fig:sen}. Even though this search can be more sensitive to light RHNs, nevertheless, its limits are not competitive but still comparable to those from dijets searches, reaching $M_{W_R} \simeq$ 3.6 TeV for $m_N \approx 5-10$ GeV. This is due to the low efficiency, $\epsilon_{\text{tot}} \simeq \mathcal{O}(10^{-3})$. With the help of more than 20 times higher luminosity, and slightly better $\epsilon_{\text{tri}}$ using softer trigger, the lower limits can be improved to $M_{W_R} \simeq$ 8.0 TeV for $m_N \approx 5-15$ GeV, at the HL-LHC with softer trigger strategy. This has surpass the current limits from $0 \nu \beta \beta$ experiments by at most 2 TeV, especially where $m_N \sim $ 10 GeV and $0 \nu \beta \beta$ becomes insensitive.

When it comes to $gg \rightarrow \Delta \rightarrow N N$ processes, due to the low trigger efficiency, we only obtain 2 signal events at most using current CMS data, failing to reach 95\% CL. Notwithstanding, relatively positive sensitivity can still be obtained using soft trigger at HL-LHC.
In the plane of ($M_{W_R}$, $m_N$), the production cross section of these processes are fixed, hence the sensitivity is only controlled by the total efficiency $\epsilon_{\text{tot}}$, which dominantly depends on the proper decay length of the $N$. This is reflected as the sensitivity roughly tracks where $0.01~\text{m} \lesssim L_0 \lesssim 1$ m, regardless of how large $M_{W_R}$ is, so it can be extended to $M_{W_R} >$ 12 TeV. When comparing the two masses of $\Delta$, the sensitivity shrinks for $m_{\Delta} = $ 150 GeV since the production cross section is more than 10 times smaller, despite the fact that the trigger efficiency is 5 times larger.

\section{Conclusion}
\label{con}
The LRSM model is one of the most attractive models to accommodate the seesaw mechanism to explain the origin of tiny neutrino masses. It offers ample opportunities to search for new phenomenology, including those in $0\nu\beta\beta$ experiments, as well as colliders. Current collider searches have been mainly focused on the prompt final states of the RHNs, while several searches for long-lived ones are also proposed and might be operated in the future at far detectors of the HL-LHC, lepton colliders, as well as FCC-hh.

In this work, we consider a more realistic analysis for long-lived final states, using the current CMS data, and future data at the HL-LHC using the updated trigger already installed. We reinterpret the existing search for displaced shower signature at the CMS muon system, to the long-lived RHNs from either
$pp \rightarrow W_R^{\pm} \rightarrow l^{\pm} N$, or $gg \rightarrow \Delta \rightarrow N N$ processes. For the former processes, we obtain the lower limits $M_{W_R} \gtrsim$ 3.6 TeV for $m_N \approx 5-10$ GeV using current CMS data. This is comparable to the existing limits from the dijet searches. And the sensitivity can be improved to $M_{W_R} \gtrsim$ 8.0 TeV for $m_N \approx 5-15$ GeV, at the HL-LHC with soft trigger, exceeding the current limits from the $0\nu \beta \beta$ experiments. 
For the latter processes, although no sensitivity at 95\% CL is obtained at current CMS, much broader region can be explored which can be extended to much larger $M_{W_R} >$ 12 TeV at the HL-LHC, giving that $m_{\Delta} \lesssim 160$ GeV.

Although the forecasting limits at the HL-LHC using displaced shower searches are still worse than the ones from looking for displaced vertex at the HL-LHC with or without far detectors, here we emphasize that our results have already considered detailed reconstruction and trigger efficiency, which is absent in the literature for the searches of displaced vertex. Taking into account these effects, the sensitivity of the displaced shower searches should be complementary to the displaced vertex searches. And performing such searches does not require further reconstruction of new detectors, or even new colliders.

\textbf{Note added.} During the end of preparation of this manuscript, Ref.~\cite{fuks2025beautifulmajoranahiggsescolliders} appeared on arXiv, in which similar processes are considered. Scalar triplets are produced in associated with either a Higgs boson or $W_R^\pm$. They further decay into long-lived RHNs, result in a long-lived lepton signature at the LHC.

\acknowledgments

W. L. is supported by National Natural Science Foundation of China (Grant No.12205153). We thank Zeren Simon Wang for useful discussions.
\bibliographystyle{JHEP}
\bibliography{main.bib}
\end{document}